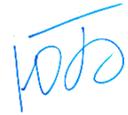

Manuscript copyright

# Barsukov Yuri Vladimirovich

# Plasma processes affecting etching and growth of nitride materials

01.04.08 – plasma physics

ABSTRACT

of thesis for the PhD degree

Saint-Petersburg

2022

Thesis is done in Peter the Great Saint-Petersburg State University

Scientific supervisor is Prof. Alexander Sergeevich Smirnov

Official opponents:

Dr. Timofeev Nickolay Alexandrovich — Prof. Saint-Petersburg State University

Dr. Schweigert Irina Vacheslalovna — Senior Researcher, Khristianovich Institute of Theoretical and Applied Mechanics

Externa Reviewer Petersburg Nuclear Physics Institute

Thesis Defense will be held on April 18 2022
at a meeting of the Dissertation Committee У.01.04.08
of Peter the Great St. Petersburg Polytechnic University.
(195251, Russia, St.Petersburg, Polytechnicheskaya, 29, building 2, room 425).

Thesis can be found in the library
and on website https://www.spbstu.ru/science/the-department-of-doctoral-studies/defences-calendar/the-degree-of-candidate-of-sciences/barsukov_yuriy_vladimirovich/
of Peter the Great St. Petersburg Polytechnic University.

The abstract is sent on ___________________
                              date

Scientific Secretary of the Dissertation Committee

Dr. Kaveeva Elisaveta Genadievna

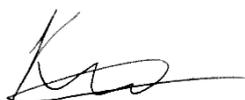

# Introduction

**Relevance of the work**

Plasma treatment is widely used in semiconductor industry during integrated circumstance fabrication. Particular, a dry isotropic etching is one of the most challenge step during FEOL (front end of line) step of nanofabrication of integrated circumstance based on CMOS technology (complementary metal oxide semiconductor) [1]. The dry etching implies interaction of the surface with neutral species, and the last ones are generated in a remote plasma source (RPS). Damage free etcher with RPS is typically used for such kind of etching, where plasma has no contact with the etched surface in order to exclude the surface damage by UV emission and ion bombardments. A sandwich-like structure with silicon oxide ($SiO_2$) and silicon nitride ($Si_3N_4$) layers is formed during one of step of 3D-NAND memory fabrication. At the next step, the $Si_3N_4$ layer is horizontally removed forming a cavities, which are in turn filled by wolfram (metallization step) [2]. A wet etching using hot phosphoric acid is used so far to remove silicon nitride with high selectivity over silicon oxide [3,4]. The $Si_3N_4/SiO_2$ selectivity is $ER(Si_3N_4)/ER(SiO_2)$ ratio, where ER is etch rate. Plasma etching is very promising technology. In future, it should fully replace wet etching in the semiconductor industry. Thus, to study a process of highly selective etching of Si3N4 by remote plasma source is relevance task for the industry.

Another important application of plasma is synthesis of new materials. For example, boron nitride BN is isoelectronic analogous of carbon, and BN is able to form similar structures as carbon [5]: fullborenes (0D), nanotubes (1D), hexagonal graphite sheets (2D), diamond like crystals (3D). The most of these structures cannot be found in nature, therefore they are produced by synthesis. Due to polarity of the B-N bond (C-C is nonpolar bond), BN compounds have unique properties. For example, BN structures are resistant to chemical attack [6] and to heating [7]. Moreover, defect free BN nanostructures have a wide band gap [8], therefore they can be used in optoelectronics [9]. BN nanotubes (BNNT) were synthesis in the labs using high temperatures synthesis. Namely, arc discharge [10,11] and ICP (inductively coupled plasma) torch [12–14] were used during the synthesis. Despite this fact, this is still no technology for a large-scale production of BNNTs. One of the main impediment to large scale production is liquid boron consumption during the synthesis. Thus, to study a process of BNNT synthesis and determine a key process of this synthesis is relevance task for the modern industry.

**Objectives and targets**

The main objective of my research is to determine an optimal conditions of gas discharges to plasma etching of silicon nitride (with high selectivity relevant to silicon oxide) and to large scale production of boron nitride nanotubes with high purity. It was suggested to determine chemical reagents to etch silicon nitride, while no silicon oxide etching. These reactants are in turn generated in the remote source of $NF_3/O_2$ and $NF_3/O_2/N_2/H_2$ plasmas. In addition, to study a mechanism of $N_2$ molecule fixation process and a precursor formation of boron nitride nanotube growth during a high temperature synthesis, and to determine key factors, which affect on the boron consumption.

It was suggested and solved the following targets:

1. Conducting of experiments towards to study a highly $Si_3N_4/SiO_2$ selective etching using remote source of $NF_3/O_2$ and $NF_3/O_2/N_2/H_2$ plasmas.

2. Using quantum chemistry methods to develop a mechanism of $Si_3N_4$ etching, determine key reagents for the selective etching, calculate rate constant of surface reactions leading to etching.

3. Based on the suggested mechanism to develop an analytical mode of the etching, which will describe a dependence of the etch rates of silicon oxide and nitride on the fluxes of reactants.

4. To make a comparison between results of analytical modelling and experimental data and prove a correctness of the model. Also, a model of $NF_3/O_2$ and $NF_3/O_2/N_2/H_2$ plasmas will be developed. In addition, densities of radicals and molecules will be measured by actinometry and mass-spectrometry.

5. To determine key reagents for the high $Si_3N_4/SiO_2$ selective etching based on the modelling and experimental data.

6. To calculate the thermodynamic potentials (Gibbs free energy) for B-N compounds. These potentials will be used to determine a composition of $B/N_2$ mixture at thermodynamic equilibrium using Gibbs free energy minimization method.

7. To make a modelling of the reactions between $N_2$ molecules and small boron clusters in gas phase and calculate rate constants of these reactions, where B-N compounds are generated ($N_2$ fixation).

8. To develop a kinetic system of equation. This system of equation will describe the process of $N_2$ fixation and formation of B-N compounds. The last ones are the precursors of boron nitride nanotube growth in the cooling gas mixture.

**Novelty**

1. It was shown that the key reagent of the etching by $NF_3/O_2$ plasma are NO (nitric oxide) and F atom. Note, that NO enhance the etching of $Si_3N_4$, while no effect on the $SiO_2$ etching. Atomic fluorine etches both silicon oxide and silicon nitride, therefore the selectivity is not high for such kind of etching.

2. A mechanism of Si3N4 etching by F atoms and NO was developed based on the results of quantum chemistry modelling. According to the suggested mechanism, new N-F bonds caused by the subsequent reaction steps are formed during the etching: F + ~Si-N-Si~ = ~Si-F + ·N-Si~; F + ·N-Si~ = F-N-Si~. Then, NO reacts with F-N bonds of the fluorinated surface resulting in enhancing migration of F atoms from the N surface atom on the neighboring Si surface atom: NO + F-N-Si~ = $N_2O$ + F-Si~. As a result, the volatile by-products of the etching are generated, such as $SiF_4$ и $N_2O$, and the etch rate increases. Suggested analytical model quantitively describes the dependence of the silicon nitride and oxide etch rates on the fluxes of F toms and NO.

3. By analogous with NO, the quantum chemistry modelling was performed to study the reactions between HF, Cl, H, Br и FNO species with F-N surface bond. According to the modelling data, these species accelerate the migration of surface F atoms also. This result is able to quantitively explain the early published experimental data.

4. For the first time, it was that the curve of $Si_3N_4/SiO_2$ selectivity in dependence of $H_2$ flow rate in the afterglow area of $NF_3/O_2/N_2$ plasma has a high and narrow peak. Namely, the selective peak is near the point, where density of F atoms equals to density of $H_2$ ($[F]≈[H_2]$).

5. It was shown that the density of the vibrational excited species HF(v) has a peak near the $[F]≈[H_2]$ point.

6. It was developed a mechanism and analytical model of the $Si_3N_4$ and $SiO_2$ etching by $NF_3/O_2/N_2/H_2$ remote plasma. According to the suggested model, the main etchants of both silicon oxide and nitride are F atoms and HF molecule in ground and vibrationally excited states. At room temperature, HF molecule at the ground state is able to etch $Si_3N_4$ and $SiO_2$ in the presence of catalysts. Particular, water molecules play a role of the catalyst in the downstream plasma. Thus, the suggested analytical mode describes the dependence of $Si_3N_4$ and $SiO_2$ etch rate on the fluxes of F, HF(v=0), HF(v=1) и $H_2O$.

7. Based the comparison of the modelling data with the experimental data, it was shown for the first time that vibrationally excited HF(v) molecules are able to initiate the etching of $Si_3N_4$, while no $SiO_2$ etching. As a result, the $Si_3N_4/SiO_2$ selectivity of this etching is very high.

8. For the first time, it was experimentally shown a kinetic isotope effect on the reactive ion etching of $Si_3N_4$: the etch rate drops by replacing $H_2$ with $D_2$ in $SF_6/H_2$ plasma.

9. For the first time, it was suggested a mechanism of $N_2$ fixation process resulting in formation of B-N compounds, which are in turn precursors of the boron nitride nanotubes growth during the high temperature synthesis. It was theoretically shown that dissociative adsorption of $N_2$ molecules on small boron clusters ($B_m$) occurs at the temperatures below boron condensation point. Eventually $n/2N_2 + B_m$ results to formation of $B_mN_n$ linear chains (with $n≤12$, $m=n+1$ or $n$).

Moreover, these $B_mN_n$ linear chains are the main components of the reaction mixture at the thermodynamic equilibrium at the temperature range from 2600 to 3000K at1 atm pressure.

10. It was shown that the process of $N_2$ fixation is the rate-limiting step, which determines the boron consumption in $B/N_2$ mixture. The boron consumption increases at high pressure and low initial boron fraction in the mixture. That make a good condition to large scale production of high-quality and purity boron nitride nanotubes.

**Practical and scientific value**

1. For the first time, it was shown that vibrationally excited molecules HF(v) initiate the highly selective etching of silicon nitride without catalyst.

2. It was suggested new reagents which are able to enhance the $Si_3N_4$ etching in the presence of F atoms, such as HF, Cl, H, Br и FNO. Moreover, these reagents should stronger enhance the etching of silicon nitride than NO.

3. For the first time, it was shown that the dissociative adsorption of $N_2$ on small boron clusters ($N_2$ fixation process) determine the boron consumption during high temperature synthesis. In addition, the system does not achieve an equilibrium during the gas cooling at a typical for the synthesis gas cooling rates. It was shown that fixation of $N_2$ increases at high pressure, low initial boron fraction and low gas cooling rate resulting in high boron consumption.

4. It was shown for the first time that B-N linear chains and monocyclic rings are the most stable B-N compounds at the narrow temperature range (~2600-3000K at 1 atm pressure).

**Thesis Statements**

1. The etch rate of the silicon nitride by the $NF_3/O_2$ remote plasma is a function of the fluxes of F atoms and NO radicals. Atomic fluorine fluorinates and etches the surface, while nitric oxide enhances the etching interacting with the fluorinated surface. A role of NO in the etching of silicon nitride by $NF_3/O_2$ remote plasma appears in the presence of F atoms only, therefore the $Si_3N_4/SiO_2$ selectivity is limited due to the F atom presence.

2. Highly $Si_3N_4/SiO_2$ selective etching by $NF_3/O_2/N_2/H_2$ remote plasma occurs near the point, where density of F atoms equals to density of $H_2$ molecules $[F] \approx [H_2]$. The key reagents of such kind of etching are F atom, HF(v=0), HF(v=1) and $H_2O$.

3. Vibrationally excited molecules HF(v) is able to initiate the highly $Si_3N_4/SiO_2$ selective etching of without a catalyst.

4. The process of $N_2$ fixation during high temperature synthesis of boron nitride nanotubes determines the rate of boron consumption. Fixation of $N_2$ molecule fixation occurs through the reactions between $N_2$ molecules and small Bm clusters.

5. The processes of $N_2$ fixation and boron consumption occurs at non-equilibrium regime at a typical gas cooling rate during high temperature synthesis.

6. The boron consumption during high temperature synthesis of boron nitride nanotubes increases at high pressure, low gas cooling rate and low initial boron fractions. s

7. B-N linear chains and monocyclic rings are thermodynamically the most stable compounds in $B/N_2$ mixture at narrow temperature range: higher the temperature suitable for the boron nitride nanotubes growth (~ 2400 K) and below boron condensation point.

**Validity of the results**

It was used the modern setups during this research. All results were reproduced. All results are in agreement with the theoretical model, do not contradict to the published data. All results were tested on the Russian and international conferences.

**Test of the research**

The results of my thesis were many times discussed reported on the seminars in the Department of Physics of Optics of Saint-Petersburg State University, Department of Quantum Chemistry of Saint-Petersburg State University, Quantum Chemistry Lab in Petersburg Institute of Nuclear Physics, Plasma Physic Department of Peter the Great Saint-Petersburg Polytechnic University, Lab of Plasma Physics and Basis of Microtechnology, "Coddan Technology" Ltd. (Russia, SPb), "Corial" company (Grenoble, France), LLC "Samsung Electronics" (South Korea), Princeton Plasma Physics Lab (USA), Applied Materials Inc. (California, USA).

This research was presented on five conferences:

- 24[th] International Symposium on Plasma Chemistry (Naples (Italy), June 9-14 2019)

- Nature – "Advances and Applications in Plasma Physics" (Saint-Petersburg (Russia), September 18-20 2019).

- International Conference PhysicA.SPb Saint-Petersburg (Russia), 2020.

- The 73[rd] Annual Gaseous Electronics Conference (virtual), 2020

- 63[rd] Annual Meeting of the APS Division of Plasma Physics (virtual), 2020

**My Publications**

[1] Yu. Barsukov, V. Volynets, S. Lee, G. Kim, B. Lee, S.K. Nam, and K. Han. Role of NO in highly selective $SiN/SiO_2$ and $SiN/Si$ etching with $NF_3/O_2$ remote plasma: Experiment and simulation // J. Vac. Sci. Technol. A. 2017, V. 35, P. 061310. DOI: 10.1116/1.5004546.

**Personal Contribution**

All presented here results were obtained by myself or with my participation. I give significant contribution in all presented here experimental measurements, treatment of the data, quantum chemistry modelling, development of the chemical mechanisms and analytical modelling.

All results of my research presented here were obtained in Samsung Electronics Ltd. and Peter the Great Saint-Petersburg Polytechnic University.

**Table of content**

Novelty, objectives and targets of this research as well as list of my publications and conferences were considered in the **Introduction.**

Literature review is presented in the **First Section**. It is considered mechanisms of the etching of silicon-based materials. Such as silicon nitride, silicon oxide and doped and undoped silicon. It is considered the etching of silicon nitride by remote $NF_3/O_2$ and $CF_4/O_2$ plasmas, as well as a role of NO in this etching. The mechanisms of silicon nitride and oxide by HF molecule with $H_2O$ and $NH_3$ catalyst were considered. The literature revie of high temperature synthesis of BNNT can be found in this section also.

The result of my research aimed to study of $Si_3N_4/SiO_2$ selective etching by $NF_3/O_2$ remote plasma is presented in **Section 2**. The experimental setup (DFE or damage free etcher) with ICP remote plasma source (RPS) is described. This setup was made in Samsung Electronics Ltd.

Optical spectra were measured in the range between 200 nm and 800 nm. Actinometry method [15–17] was used to measure the densities of F and O atoms, therefore small amount of Ar was injected in the plasma zone. Due to the fact, only limited amount of species can be measured a plasma modelling was performed using Global Kin code [18,19]. The measured and calculated densities of F toms and NO radicals are presented in Fig. 1.

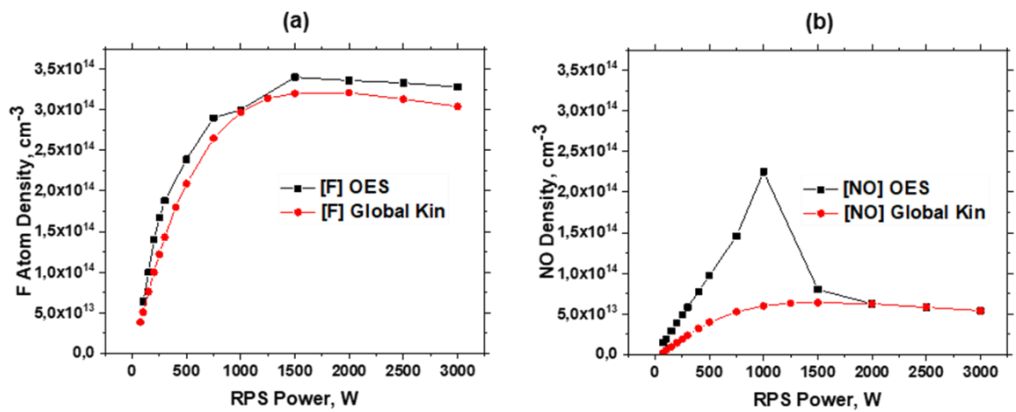

*Fig 1. Dependence of concentrations [F] (a) и [NO] (b) on plasma power.*

The calculation very accurately reproduces the density of F atoms as opposite to NO. Moreover, the calculation showed the monotonic growth of the NO density with plasma power, while measured emission of NO radical has a peak near 1000 W power. Note, the measured density of NO was used in the analytical modelling.

***Table 1.*** *List of reaction, which were considered in the model of $Si_3N_4$ etching by F and NO.*

| $R_i$ | Поверхностная реакция | $\gamma_i$ |
|---|---|---|
| **R1** | F + θ1 → $SiF_4$ + θ2 | $1.5 \times 10^{-3}$ |
| **R2** | F + θ2 → θ3 | 1.0 |
| **R3** | F + θ3 + bulk → $SiF_4$ + $N_2$ + θ1 | $1.2 \times 10^{-4}$ |
| **R4** | NO + θ3 + bulk → $SiF_4$ + $N_2O$ + θ1 | $1.8 \times 10^{-4}$ |

Based on quantum chemistry calculations, the mechanism of $Si_3N_4$ etching by F atoms and NO were suggested. According to this mechanism, NO radicals promote the migration of F atoms from the surface N atoms on the neighboring Si atoms. In other words, NO decreases the barrier of this migration resulting in enhanced silicon nitride etching.

The analytical model of $Si_3N_4$ and $SiO_2$ etching were suggested. This model described the dependence of the etch rates on the fluxes of F and NO.

The result of analytical model as well as experimental data are presented in Fig. 2. That model implies zero etch rate in the absence of atomic fluorine. In other words, NO can enhance the etching, NO cannot etch without atomic fluorine.

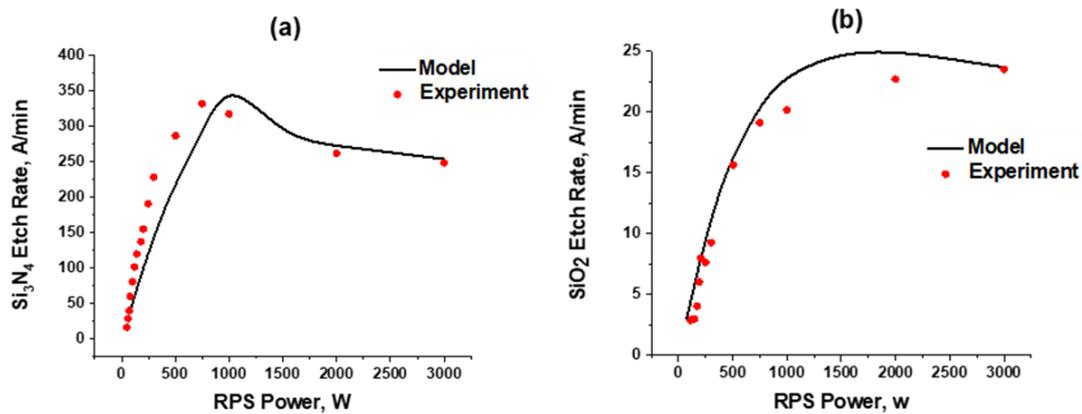

***Рисунок 2.*** *Calculated and measured etch rates of silicon nitride (a) and oxide (b) in dependence of the $NF_3/O_2$ plasma power.*

It is known from previous publications [20] that NO enhances the etching of silicon nitride. According to the suggested here mechanism of the etching F atom breaks Si-N bond resulting in formation of new Si-F bond and free bond on the N atom. This free bond is closed by the second F atom resulting in F-N bond formation. Then, NO reacts with the F-N bond and, as a result, F atom migrate from F-N bond on the neighboring Si atom and $N_2O$ molecule is formed. Note, eventually the final products of these reaction are volatile by-products, such as $SiF_4$, $N_2$ и $N_2O$.

These by-products were experimentally detected during the etching of silicon nitride [21]. Thus, atomic fluorine is essential for sun kind of etching, therefore $Si_3N_4/SiO_2$ is limited.

**Section 3** is aimed to quantum chemistry modelling of the F atom migration from the F-N in $SiF_3$-NF-$SiF_2$-NF-$SiF_3$ cluster under treatment by NO, HF, Cl, H, Br and FNO. According to the suggested mechanism, this migration is a rate-limiting step of the etching. It was shown that NO, HF, Cl, H, Br и FNO accelerates of the migration. Moreover, HF, Cl, H, Br и FNO accelerate faster this migration than NO. This fact is able to quantitively explain some published experimental data. As it was mentioned above the presence of F atoms significantly lowers the $Si_3N_4/SiO_2$ selectivity, therefore it was suggested to add $H_2$ in the $NF_3/O_2$ mixture. Gradual addition of H2 decreases density of F atoms through F + $H_2$ → HF + H reaction resulting in high $Si_3N_4/SiO_2$ selectivity.

**Section 3** is aimed to study highly selective $Si_3N_4/SiO_2$ etching by $NF_3/O_2/N_2/H_2$ plasma. It was shown that $Si_3N_4/SiO_2$ selectivity dramatically increases near the point, where [F]≈[$H_2$]. Note, that the densities of atomic fluorine and $H_2$ molecule are coupled by the following reaction

F + $H_2$ → HF(v) + H.

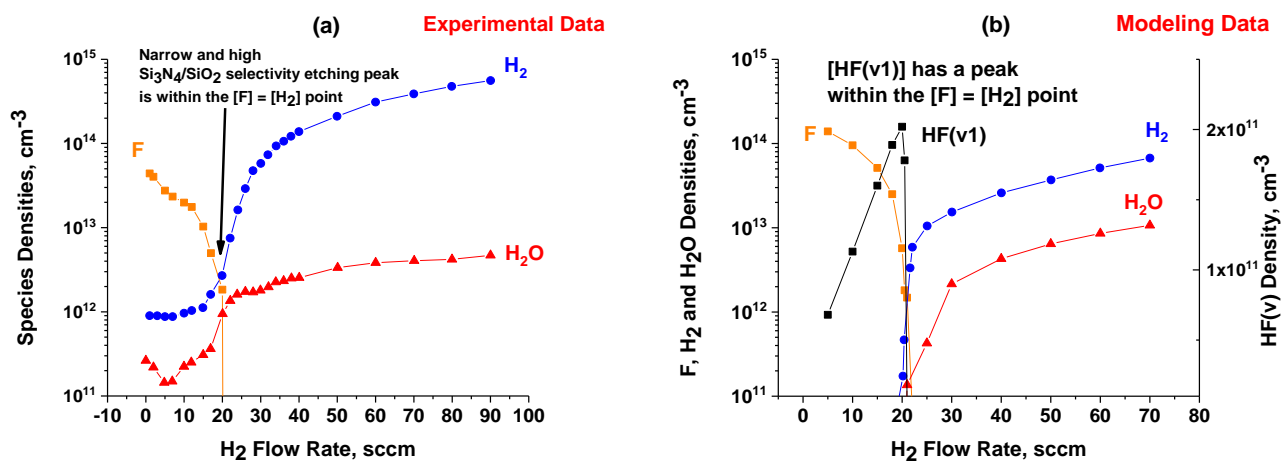

***Рисунок 3.*** *Measured (a) and calculated (b) densities [F], [$H_2$], [$H_2O$] and [HF(v1)] in the process chamber in dependence on $H_2$ flow rate.*

Densities of F and $H_2$ were measured using actinometry and mass-spectrometry methods. As it is shown in Fig. 3a both F and $H_2$ densities ([F] and [$H_2$]) strongly depends on $H_2$ flow rate. At low $H_2$ flow rate the density of F atoms monotonically decreases with $H_2$ flow rate increasing. The decline of F atom density becomes stronger at $H_2$ flow rate higher than 20 sccm. Then [F] achieves a plateau at $H_2$ flow rate in the range 22 - 90 sccm. The density of $H_2$, which was measured using mass-spectrometry, has the different trend. Indeed, at low $H_2$ flow rate (0 - 20 sccm) the density of $H_2$ slowly increases, then this growth dramatically accelerates at the ~ 20 sccm $H_2$ flow rate.

Sharp decline of [F] and sharp growth of [$H_2$] occurs near the [F]≈[$H_2$] point. Optical spectrum was measured in the 200 nm – 800 nm range, which does not cover HF(v) emission (2.5 μm). Therefore, [HF(v1)] was calculated using plasma modelling code. As it is shown in Fig. 3b, the densities of atomic fluorine, molecular hydrogen and water agreed with the measured densities well. Note, the density of HF(v1) molecular at fist vibration excitation has a high peak in the same area (it is near the [F]≈[$H_2$] point), where $Si_3N_4$/$SiO_2$ selectivity has a high and narrow peak. According to the modelling data, the density of HF(v2) is near $2 \times 10^{11}$ cm$^{-3}$. According to the analytical modelling, this density number is enough to support the etch rate of $Si_3N_4$ on a significant level near the [F]≈[$H_2$] point. At the same time, SiO2 etch rate has a minimum value near the [F]≈[$H_2$] point, because the densities of both F and $H_2O$ are very low near this point. Indeed, it is known that $H_2O$ catalyzes the $SiO_2$ etching by HF molecule at the ground state. It is known, that dry HF gas in the absence of water vapors begins to etch $SiO_2$ at T>1000 K [22]. Water is catalysis of the etching [23] and the presence of water is essential for the etching by HF. In other words, near the [F]≈[$H_2$] point $SiO_2$ etch rate has a minimum value as opposite to $Si_3N_4$ etch rate, because $Si_3N_4$ is etched by vibrationally excited HF(v) molecules.

The surface reactions which were used in the analytical model of the etching is shown in Table 2. The following surface sites were introduced: $\theta_1$ – initial surface site of $Si_3N_4$ or $SiO_2$, $\theta_2$ – surface sites with adsorbed HF molecule at the ground state HF(v=0), $\theta_3$ – surface site with two adsorbed HF molecules HF(v0) + HF(vs1) one of which is vibrationally excited molecule, $\theta_4$ – surface site with HF molecule at the ground state and water molecule HF(v=0) + $H_2O$.

We assumed that the etching occurs at the steady state conditions and the surface composition is not a function of time. Moreover, we assumed that the surface reactions do not significantly change the fluxes of the reactant near the surface. Molecules of HF(v0), HF(v1) and $H_2O$, which are generate in the downstream plasma, are adsorbed on the $\theta_1$ surface sites resulting in formation of $\theta_2$, $\theta_3$ или $\theta_4$ sites (R1, R2 and R5 reactions in Table 2). Desorption of these species from the surface leas to the formation of $\theta_1$ sites (R3 and R4 reactions in Table 2). Reactions of the etching leads to formation of volatile by-products ($SiF_4$, $NH_3$, HF and $H_2O$), which is denoted as EP or etch products (R6, R7 and R8 reactions). Only HF(v1) molecules at the fist vibrational excitation was taking into account. These vibrationally excited HF(v1) molecules are adsorbed on $\theta_2$ sites (R2 reaction), and the etching occurs through R7 reaction.

***Table 2.*** *List of the reactions and rate constants, which was used in the analytical of Si3N4 and SiO2 etching by F, HF, H₂O и HF(v1).*

|  | Поверхностные реакции | $\gamma$ | | A (1/s) | | $E_a/R$ (K) | |
|---|---|---|---|---|---|---|---|
|  |  | $Si_3N_4$ | $SiO_2$ | $Si_3N_4$ | $SiO_2$ | $Si_3N_4$ | $SiO_2$ |
| **R1** | HF + $\theta_1$ → $\theta_2$ | 1 | 1 | - | - | - | - |
| **R2** | HF(v=1) + $\theta_2$ → $\theta_3$ | 1 | 1 | - | - | - | - |
| **R3** | $\theta_2$ → HF + $\theta_1$ | - | - | $5.00 \times 10^7$ | $9.00 \times 10^9$ | 2044 | 1996 |
| **R4** | $\theta_3$ → 2HF + $\theta_1$ | - | - | $2.10 \times 10^9$ | $8.00 \times 10^{10}$ | 3612 | 0 |
| **R5** | H₂O + $\theta_2$ → $\theta_4$ | 1 | 1 | - | - | - | - |
| **R6** | $\theta_4$ → EP + $\theta_1$ | - | - | $1.18 \times 10^{10}$ | $1.46 \times 10^{13}$ | 7862 | 7399 |
| **R7** | $\theta_3$ → EP + $\theta_1$ | - | - | $4.02 \times 10^{11}$ | $6.55 \times 10^8$ | 1756 | 8061 |
| **R8** | F + $\theta_1$ → EP + $\theta_1$ | 0.00001 | 0.0000195 | - | - | - | - |

The rate constant of R3, R4, R6 и R7 reactions are in Arrhenius form:

$$k_i = A_i \exp\left(-\frac{E_{ai}}{RT}\right). \quad (3)$$

Activation energies ($E_a$) were calculated as an energy difference between reactant complex and transition state. Transition state theory [24] and Eyring equation [25] were used to calculate a pre-exponential factors $A_i$ for R6 и R7 reactions,

$$A_i = \frac{k_B T}{h} \times \left(\frac{Z^{TS}_{vib}}{Z^{Re}_{vib}}\right), \quad (4)$$

where $k_B$ и h are Boltzmann and Planck constants, $Z^{TS}_{vib}$ and $Z^{Re}_{vib}$ – vibrational partition functions of transition start and reactant complex, respectively. The partition functions were calculated using Gaussian 09 software package [26].

Reagents in the R6 and R7 reactions are adsorbed molecules: HF + H₂O и HF + HF(v1). Molecule of HF reacts with both $Si_3N_4$ and $SiO_2$ at room temperature only in the presence of the catalyst (water in our case). Internal vibrational energy of HF(v1) molecule is used to overcome the barrier of the surface reaction leading to the etching of $Si_3N_4$. It is known that the vibrational quantum energy of O-H and F-H bonds are very close to each other, therefore HF(v1) should be quasi-resonantly quenched in a collision with hydroxyl groups of oxide surface. Thus, the residence time of the vibrationally excited HF(v) molecules is a function of the average time of the etching reactions and quenching. Due to the fact that the transition of vibrational quantum from HF(v) molecule on the hydroxyl groups is quasi-resonance process, molecules of HF(v1) are quenched on the oxide surface without etching.

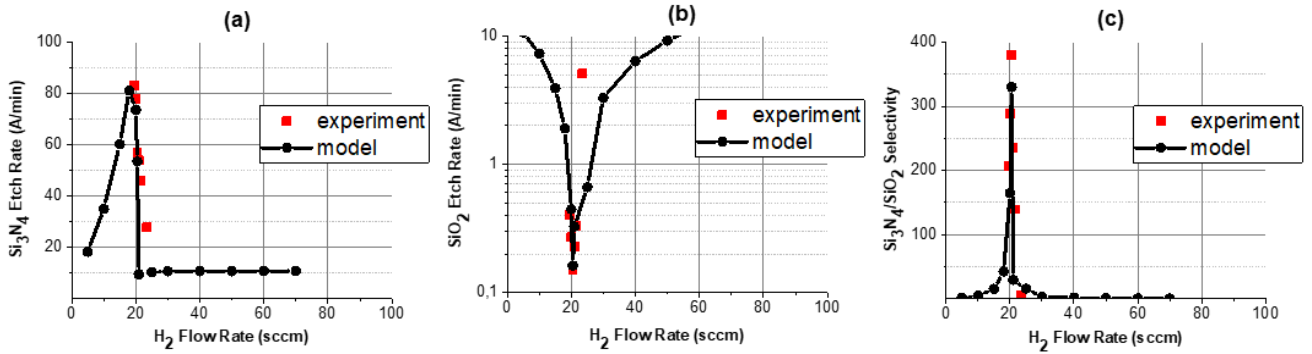

*Figure 4. Measured and calculated etch rates of $Si_3N_4$ and $SiO_2$ as well as $Si_3N_4/SiO_2$ selectivity in dependence of $H_2$ flow rate: (a) etch rate of $Si_3N_4$, (b) etch rate of $SiO_2$ and (c) $Si_3N_4/SiO_2$ selectivity.*

The data of the etching and the selectivity data, which were calculated using the analytical model of the etching as well as experimental data, are presented in Fig. 4. The calculated data are correlate with the experimental data well. The etch rate of $Si_3N_4$ and $Si_3N_4/SiO_2$ selectivity peaks near 20 sccm $H_2$.

The contribution of F, HF + $H_2O$ and HF + HF(v1) to the total etch rate of $Si_3N_4$ and $SiO_2$ in dependence of $H_2$ flow rate is shown in Fig. 5. The etch rate by F atoms decreasing with $H_2$ flow rate, because the density of F decreases with $H_2$ flow rate. The contribution of HF(v1) to the etching of $Si_3N_4$ is determined by the flux of HF(v1). The flux of HF(v1) peaks near the [F]≈[$H_2$] point. The water production is slow at [$H_2$]<[F], and it dramatically increases at [$H_2$]>[F]. As a result, the etching by HF + $H_2O$ increase at [$H_2$]>[F], when the flow rate of $H_2$ is higher than 20 sccm. Thus, the etch rate of silicon nitride has no minimum near the [F]≈[$H_2$] point as opposite to silicon oxide, since HF(v1) contribution to the etching is significant for silicon nitride only.

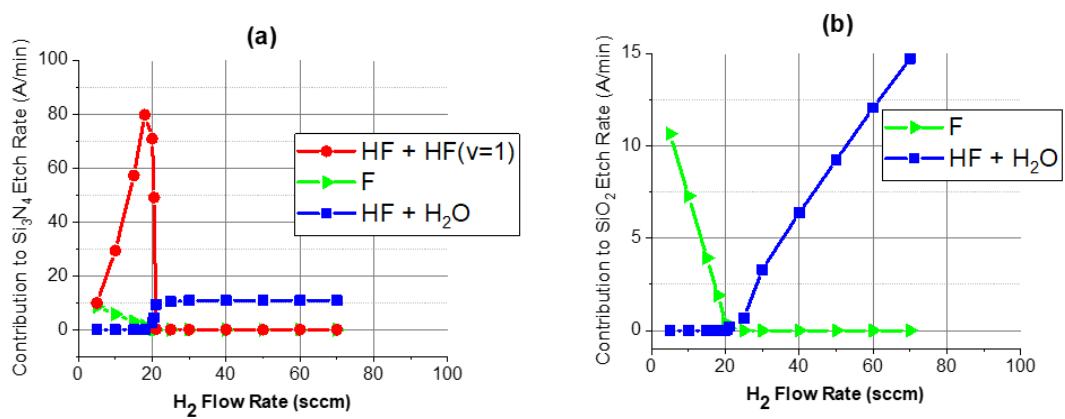

*Figure 5. Contribution to the etching of: (a) $Si_3N_4$ and (b) $SiO_2$.*

**Section 5** is aimed to study of reactive of ion etching of silicon nitride and oxide using $SF_6/H_2$ and $SF_6/D_2$ plasmas. It was shown a kinetic isotope effect on the etching of silicon nitride only. It was used a handmade setup with CCP reactor at 40.68 MHz. Plasma interacted with the samples, the last one was located between two electrodes. Ar was injected (1 sccm) to use actinometry. $H_2$ and $D_2$ flow rates were varied from 0 to 10 sccm, other conditions were fixed: 3 sccm $SF_6$, 300 W, 60 mTorr.

It is shown in Fig. 6a the dependence of density of atomic fluorine on the flow rates of $H_2$ and $D_2$. The density of $S_2$, which is generated at relatively high level, cannot be accurately measured using actinometry, because the threshold energy of the cross section of the excitation of $S_2$ and Ar are significantly differ. Therefore, the intensity of $B^3\Sigma^-_u(v0) \rightarrow X^3\Sigma^-_u(v9)$ transition (283 nm) was normalized on the intensity of Ar transition (750 nm). It is assumed that $I(S_2)/I(Ar)$ correlates with real density of $S_2$. The $I(S_2)/I(Ar)$ ratio increases with hydrogen flow rates for both $SF_6/H_2$ and $SF_6/D_2$ mixtures, achieving a plateau at 6 sccm (Fig. 6b).

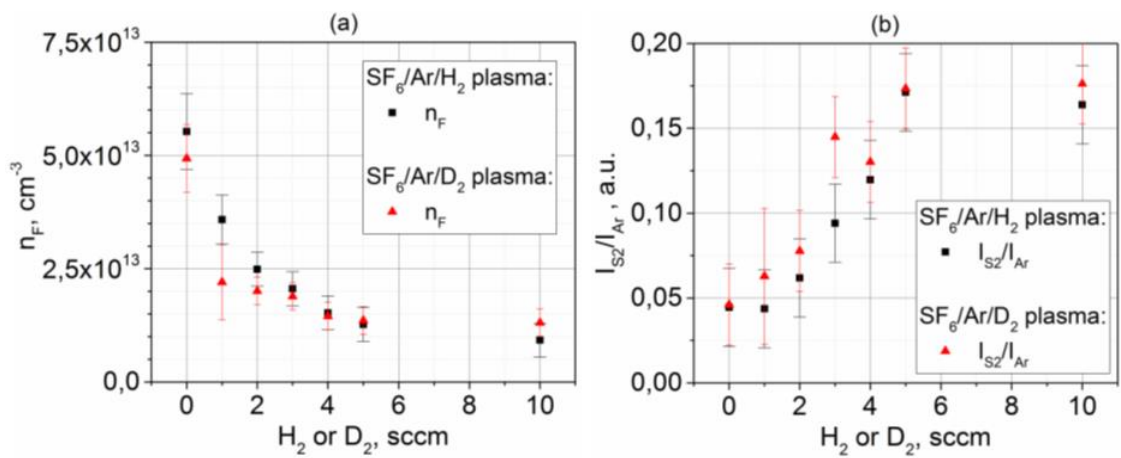

*Figure 6. Dependence of the atomic fluorine density (a) and $I(S_2)/I(Ar)$ ratio on the hydrogen flow rate ($H_2$ and $D_2$).*

The dependence of the etch rate on the flow rates of hydrogen is shown in Fig. 7. According to this data the silicon nitride is etched faster even in pure $SF_6$ plasma. Small addition of $H_2$ to $SF_6$ discharge tends to faster etching of $Si_3N_4$. (the etch rate peaks at 2 sccm $H_2$). On the other hand, the etch rate of silicon nitride monotonically decreases with $D_2$ flow rate. Note that the F atom density monotonically decreases with hydrogen flow rate (Fig. 6a), therefore the etch rate peak in $SF_6/H_2$ plasma can be explained by the fact that vibrationally excited HF(v) molecules give contribution to the etching of silicon nitride. Note, the etch rate of $Si_3N_4$ monotonically with $D_2$ flow rate decreases in $SF_6/D_2$ plasma. This etch rate decline correlates with the atomic fluorine density dropping. Indeed, the vibrational quantum of DF is significantly lower than HF (0.3 eV vs 0.5 eV). Thus, it can be assumed that DF(v) has no enough energy to give visible contribution to the etching of silicon nitride. The observed here isotope effect is not very high, because the density of atomic fluorine is very high (~ $10^{13}$ cm$^{-3}$ in Fig. 6a), therefore atomic fluorine gives the main contribution to the etching rather than vibrationally excited HF(v) molecules. Moreover, ion bombardments clean the $Si_3N_4$ surface from the HF molecules at the ground and vibrationally excited states.

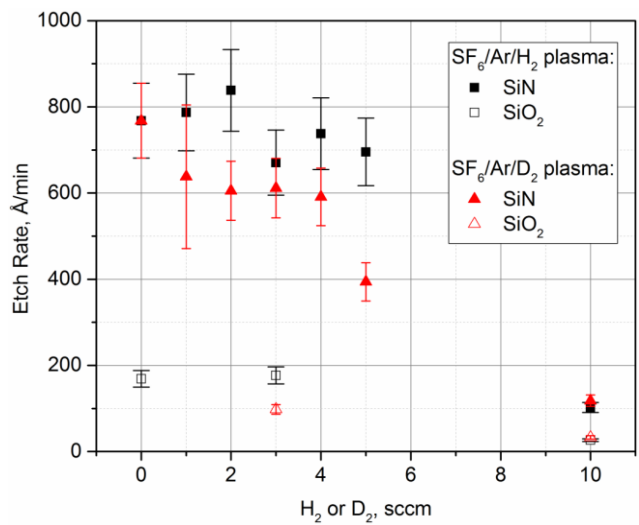

*Figure 7. Dependence of $Si_3N_4$ and $SiO_2$ etch rates in $SF_6/H_2$ and $SF_6/D_2$ plasmas on hydrogen flow rate.*

**Section 6** is aimed to study (using integrated modelling) of mechanism precursor formation of boron nitride nanotubes growth during high temperature synthesis. First, compositions of $B/N_2$ mixture at thermodynamic equilibrium at various pressures and temperatures were calculated using Gibbs free energy minimization method (thermodynamic approach). Then system of kinetic equation, which describes a dissociative adsorption of $N_2$ molecules on small boron clusters ($N_2$ fixation), was numerically solved; gas cooling rate was taken into account (kinetic approach). The kinetic approach was used to check the assumption that thermodynamic equilibrium is valid during the $N_2$ fixation process.

The thermodynamic approach shoed (Fig. 8a and 8b) that all liquid boron is consumed and converted into $B_{13}N_{12}$ chains at the temperature higher than temperature suitable for BNNT growth (~ 2300 K). The liquid boron consumption depends on pressure. Indeed, all liquid boron is consumed at T<2800 K (1 atm) and at T<3600 K (10 atm).

Addition of hydrogen to the reaction mixture leads to the production of HBNH iminoborane (Fig. 8c). The density of HBNH similarly changes with temperature as the density of $B_mN_n$ chains. Thus, it can be assumed that HBNH molecule play a crucial role during the BNNT in a hydrogen contained mixtures.

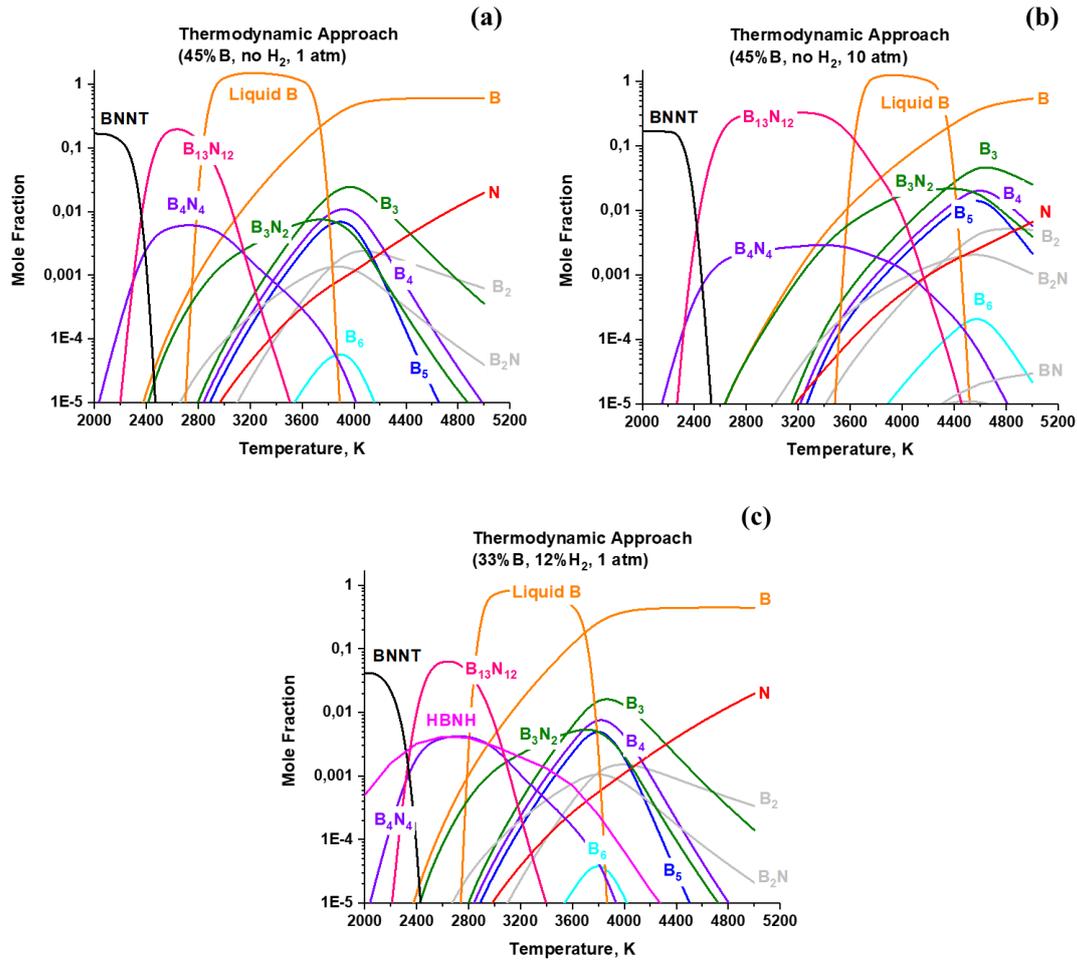

*Figure 8.* *Composition of B/$N_2$ mixture at thermodynamic equilibrium at 1 atm.(a), at 10 atm.(b), for B/$N_2$/$H_2$ mixture at 1 atm.(c).*

It is shown in Fig. 9 that the mole fractions of $B_4N_4$ and $B_5N_4$ significantly deviates from the equilibrium values at atmosphere pressure and higher. Thus, the kinetic approach as opposite to the thermodynamic approach showed that not all liquid boron is consumed at typical $\dot{T}_0=10^5$ K/s gas cooling rate [13] and 1 atm pressure. High pressure shifts the system towards the equilibrium, and all liquid boron is consumed at 10 atm through the reactions of dissociative adsorption of $N_2$ on small boron clusters (Fig. 9c).

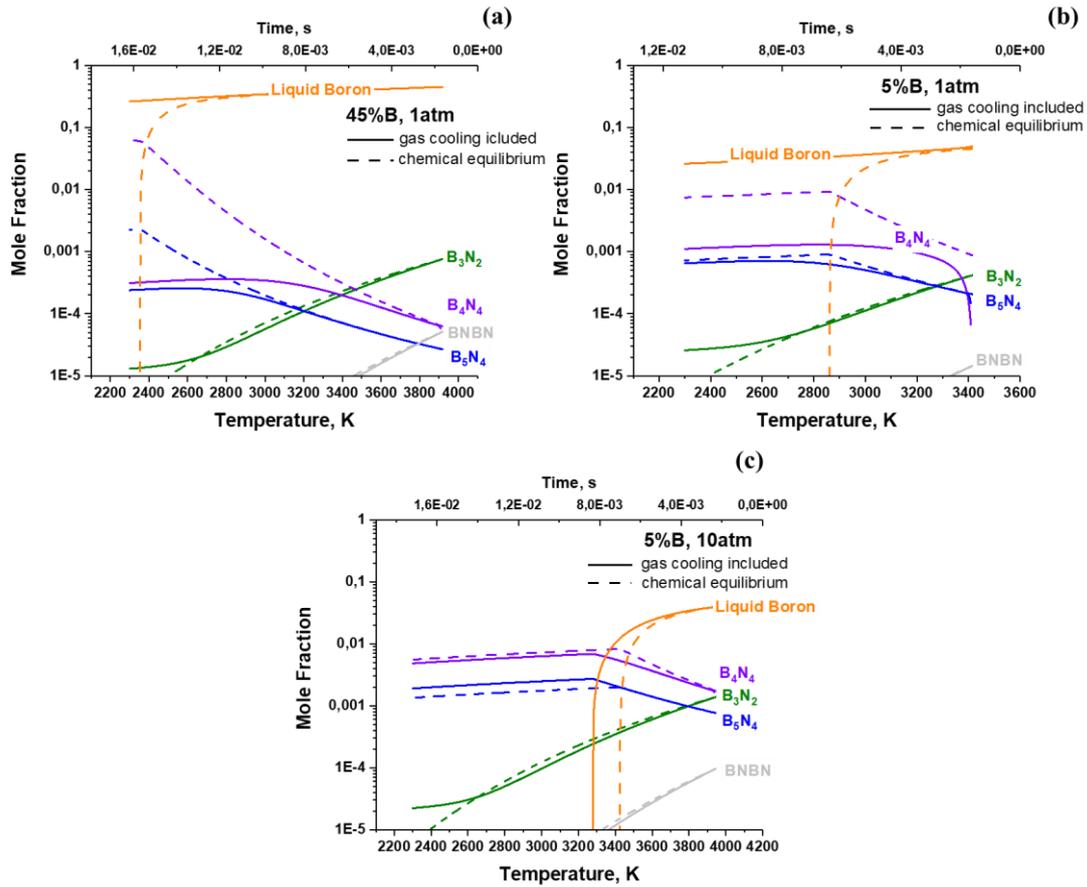

**Figure 9.** *Results of kinetic modelling of the process of $N_2$ fixation at $\dot{T}_0=10^5$ K/s gas cooling rate and various conditions: 45%B and 1 atm (a), 5%B and 1 atm, 5%B and 10 atm.*

## Conclusion

The main goal of this research was to determine the key reagents during the highly $Si_3N_4/SiO_2$ selective etching. The experiments were conducted, where both $Si_3N_4$ and $SiO_2$ samples were etched by $NF_3/O_2$ and $NF_3/O_2/N_2/H_2$ plasmas. The sources of the plasmas were removed from the etched sample to exclude a damaging from UV emission and ion bombardment. Optical emission spectroscopy and mass-spectroscopy were used during the etching. The reaction mechanisms were studied using quantum chemistry methods. It was suggested analytical models, which quantitively describe dependence of $Si_3N_4$ and $SiO_2$ etch rate on the fluxes of key reactants in $NF_3/O_2$ and $NF_3/O_2/N_2/H_2$ downstream plasmas. The densities of the kye reagents were measured and calculated using plasma simulation. Thus, the $Si_3N_4$ etch rate curve in $NF_3/O_2$ mixture has a peak, where NO density peaks. The high and narrow $Si_3N_4/SiO_2$ selectivity peaks, which appears in $NF_3/O_2/N_2/H_2$, correlates with high and narrow density peak of vibration excited HF(v1) molecule.

Also, this research was aimed to study a mechanism of precursor formation of boron nitride nanotubes growth during high temperature synthesis. It was shown that boron consumption (it is the main impediment to large scale production) occurs through the reactions of $N_2$ dissociative adsorption on small boron clusters ($N_2$ fixation) resulting in generation of $B_4N_4$ and $B_5N_4$ chains. The liquid boron is only source of the small boron clusters. A subsequent formation of longer chains occurs via collisions of $B_4N_4$ and $B_5N_4$ with each other. It was also shown that slow gas cooling rate and high pressure enhance liquid boron consumption during the synthesis, creating a good condition to large scale production of high purity and quality BNNT.

The main goal of this research was achieved. The following results can be highlighted:

1. Experimental data of highly $Si_3N_4/SiO_2$ selective etching by $NF_3/O_2/N_2/H_2$ plasma. was obtained. The $Si_3N_4/SiO_2$ selective is the highest among the published data (~ 380).

2. The models of $NF_3/O_2$ and $NF_3/O_2/N_2/H_2$ plasmas were developed. Calculated densities of some reagents agree with the measured densities well.

3. It was suggested the analytical models of $Si_3N_4$ and $SiO_2$ etching by remote $NF_3/O_2$ and $NF_3/O_2/N_2/H_2$ plasma sources. This models accurately reproduce measured etch rates.

4. It was shown for the first time the vibrationally excited HF(v) molecules at low density of F atoms initiate $Si_3N_4$ etching, while HF(v) does not affect on $SiO_2$ etching.

5. The detailed mechanism of $N_2$ fixation during high temperature synthesis was suggested.

6. It was shown for the first time, that $N_2$ fixation and boron consumption processes occurs at non-equilibrium regime and they depend on gas cooling rate, pressure and initial boron fraction.

7. It was shown for the first time that BN chains are the most stable compounds at the narrow temperature range. That fact complicates the synthesis of BN chains.

It can be concluded that all results are reasonable and responded to the goals of this research. The results of my research can be used to develop recipes for plasma etching in the semiconductor industry and to design a production line to large scale synthesis of boron nitride nanotubes.

It should be note that the $Si_3N_4/SiO_2$ selectivity by vibrationally excited HF(v) molecules can be higher than here. But it is required the additional research and experiments.